\documentstyle[prb,aps]{revtex}
\begin{document}
\draft
\title{Electronic Raman scattering in the single-CuO$_2$ layered
superconductor Tl$_2$Ba$_2$CuO$_{6+\delta}$}

\author{L.V. Gasparov}
\address{2. Physikalisches Institut, RWTH-Aachen, 52056 Aachen, Germany\\
Institute for Solid State Physics, 142432, Chernogolovka,
Moscow district, Russia}

\author{P. Lemmens}
\address{2. Physikalisches Institut, RWTH-Aachen, 52056 Aachen, Germany}

\author{M. Brinkmann}
\address{2. Physikalisches Institut, RWTH-Aachen, 52056 Aachen, Germany}

\author{N.N. Kolesnikov}
\address{Institute for Solid State Physics 142432, Chernogolovka,
Moscow district, Russia} 

\author{G. G\"untherodt}
\address{ 2. Physikalisches Institut, RWTH-Aachen, 52056 Aachen, Germany}
\date{\today}
\maketitle

\begin{abstract}

Electronic Raman scattering in Tl$_2$Ba$_2$CuO$_{6+\delta}$ (Tl-2201) has
been investigated in order to test whether the scattering cross-section in high
temperature superconductors depends on the number of CuO$_2$-planes, i.e.
sheets or specific details of the
Fermi surface. The polarized Raman spectra have been  measured in different
scattering geometries for temperatures above and below T$_c$. The spectral
features of Tl-2201 with one CuO$_2$-plane per unit cell are found to be
similar to Tl$_2$Ba$_2$Ca$_2$Cu$_3$O$_{10}$ with three CuO$_2$ planes
and those of other high temperature superconductors with several
CuO$_2$-planes per unit cell.
The peak in the B$_{1g}$ symmetry component of the scattering intensity is
found at 460cm$^{-1}$ (T$_c$=90K), or 430cm$^{-1}$ (T$_c$=80K). The B$_{1g}$
peak positions scale with T$_c$, and correspond to
$2\Delta /k_BT_c=7.6\pm 0.4$. The temperature dependence of the B$_{1g}$
scattering component of Tl-2201 (T$_c$=80K and 90K) reveals a deviation
from BCS behavior. The experimental data are in qualitative agreement with
the calculations of Devereaux and Einzel based on the $d_{x^2-y^2}$-wave
symmetry of the order parameter used in the description of the Raman
scattering cross section. 
\end{abstract}

\section{Introduction}

Since the discovery of high temperature superconductors (HTSC), the pairing
mechanism and the symmetry of the order parameter in these compounds are 
key questions at stake \cite{Dyn94,Sch94}. There are several experimental
techniques which are able to address this problem. The experiments on
quasiparticle tunneling \cite{Val91}, the linear temperature dependence of
the penetration depth \cite{Har93}, the NMR and NQR measurements 
\cite{Kit93,Sli93}, and angular resolved photoemission experiments in
Bi$_2$Sr$_2$CaCu$_2$O$_8$ \cite{She93,Ma} have yielded results consistent
with $d$-wave pairing. On the other hand, quasiparticle tunneling,
the exponential temperature dependence of the penetration depth, as well
as the measurements of the electronic Raman scattering in
Nd$_{2-x}$Ce$_x$CuO$_4$
are consistent with s-wave pairing \cite{Tra91,Mao94,Hac}.
The measurements of the magnetic field dependence of the dc-SQUID
(YBaCuO-Au-Pb arrangement) \cite{Woh94} clearly indicated d-wave behavior,
while the experiments on single Josephson-junction Pb-YBCO \cite{Sun94}
showed s-type behavior. So while the experimental evidence in favor of
d-wave symmetry of the order parameter continuously grows, there
is still no final consensus about it.

Raman scattering is a powerful tool to address the problem of the symmetry of
the order parameter. It allows to probe the symmetry of the scattering
tensor by simply choosing different  polarization directions of the incident
and scattered light.
From the investigations of the Raman scattering in conventional
superconductors it is known that the superconducting transition manifests
itself in a renormalization of the electronic Raman scattering intensity
below T$_c$. It was found for Nb$_3$Sn and V$_3$Si \cite{Hac83,Die83} that
normalized Raman spectra of these compounds show for temperatures below T$_c$
a peak associated with the pair breaking process at the energy 2$\Delta$,
together with a strong decrease of the scattering intensity at frequencies
lower than 2$\Delta$. In high temperature superconductors, the first
measurements of electronic Raman scattering were reported in
Refs.\onlinecite{Lyo87,Baz87,Hac88,Coo88}. But in this case the behavior of
the electronic scattering differs from that in conventional superconductors:
A pair breaking peak develops in the spectra below T$_c$, but the scattering
intensity  at frequencies below 2$\Delta$ does not show the usual sharp
decrease. Instead, a monotonic decrease toward zero frequency is found.
Moreover, for different symmetry components (A$_{1g}$, B$_{1g}$ and B$_{2g}$)
 the renormalization of the scattering intensity for T$<$T$_c$ is different
 and they exhibit peaks at different frequencies
\cite{Hac88,Coo88,Hof94,Sta92,Hof95,Nem93,Che92,Che94,Dev94,Dev95,Kra94,Kra95}.
These facts have been explained by Devereaux et al. \cite{Dev94,Dev95} in
terms of d$_{x^2-y^2}$-wave pairing. Their calculations of the scattering
cross-section have been performed for a cylindrical single-sheeted Fermi
surface in the framework of the kinetic equation approach. The symmetry of the
crystal was taken into account through calculating the Raman vertex,
which was expanded in terms of a complete set of crystal harmonics defined
on the Fermi surface. It was found that nontrivial coupling between the
Raman vertex and an assumed strongly anisotropic energy gap leads to the
strong symmetry dependence of the scattering intensity. The calculations
\cite{Dev94,Dev95} predict specific symmetry dependencies of the low
frequency scattering as well as the peak positions for the different
symmetry components of the electronic Raman scattering at temperatures
below T$_c$. The A$_{1g}$ peak position is sensitive to the parameters of
the model calculation. It will appear below
B$_{1g}$ peak position while with some parameters it may also appear at the
B$_{1g}$ peak position. Nevertheless there is one set of parameters
which can perfectly reproduce the experimental data\cite{Priv}.
 Later on this model was criticized by Krantz and Cardona
\cite{Kra94,Kra95}. Their calculations \cite{Kra95} are based on the general
description of the Raman scattering cross-section through the inverse effective
mass tensor. In case of the multisheeted Fermi surface,(e.g. several CuO$_2$
planes per unit cell in HTSC) polarization dependent Raman efficiencies are
determined by the averages of the corresponding effective mass fluctuations.
The authors of Ref. \onlinecite{Kra95} used the effective masses from LDA band
structure calculations for YBa$_2$Cu$_3$O$_7$ to determine the Raman scattering
cross-section. They found that it contradicts the experimental results if one
uses only d-wave pairing for a multisheeted Fermi surface of YBa$_2$Cu$_3$O$_7$.
An explanation was given by assuming different types of the order parameter on
different sheets of the Fermi surface. For  a single-sheeted Fermi
surface (i.e. one CuO$_2$-plane per unit cell) no mass fluctuations should
occur. Therefore in the framework of the effective mass fluctuation approach,
the A$_{1g}$ scattering component will be nearly totally screened and should
peak at the same position where the  B$_{1g}$ scattering component
(2$\Delta_{max}$).
Therefore straihgtforward measurements of the electronic Raman scattering in
single-CuO$_2$ layered  high-temperature superconductors (Tl-2201, La-214,
Bi-2201, (Nd,Ce)-214) should clarifay this controversial point.

Tl-2201 has the highest T$_c$ (up to 110K) \cite{Kol92} among the above
mentioned single-CuO$_2$ layered compounds. Therefore all effects due to the
gap opening are expected in the range of 300-600cm$^{-1}$, and they should
not be obscured due to the Rayleigh scattering at small wavenumbers. Nevertheless,
Raman  measurements in only one pure scattering geometry (B$_{1g}$) are known
\cite{Nem93,Blu94} for this compound, which showed \cite{Nem93} becides a
T$_c$=80K two additional transitions at 100 and 125K, which can be indicative
of the Tl-2212 and Tl-2223 phases.

These facts lead us to reinvestigate the electronic Raman scattering in the
Tl-based high temperature superconductor Tl-2201 (with different oxygen
content) with one CuO$_2$-plane per unit cell. The comparison with the
results of electronic Raman scattering experiments reported for the high
temperature superconductors with several CuO$_2$-planes should clarify 
whether the multiband scattering is indeed important. We should mention that
similar experiments on the single layered compound (La-214) were already
carried out \cite{Che94}. Nevertheless, in the framework of comparison of the
compounds with different number of CuO$_2$-planes the measurements on Tl-2201
are more favourable due to its high T$_c$.

\section{EXPERIMENT}

The investigated single crystals of Tl$_2$Ba$_2$CuO$_{6+\delta}$ (Tl-2201)
had the shape of rectangular platelets with the size of approximately
2x2x0.2 mm$^3$.
The two crystals investigated were characterized by a SQUID magnetometer.
T$_c$ was found to be 90$\pm$3 K and 80$\pm$5 K. The crystals are slightly
underdoped. It is known \cite{Shi92} that
differences in T$_c$ in Tl-2201 compound originate from different oxygen
concentrations. These crystals can  be over- as well as underdoped. The
heavily oxygen doped crystals show a metallic type of  conductivity
\cite{Shi92}
and do not show a superconducting transition. The orientation of the
tetragonal crystals was controlled by X-ray diffraction.

Raman measurements were performed on "as grown" surfaces of the freshly
prepared crystals. This is very important, because the crystal surface of
Tl-based  superconductors as well as of all high temperature superconductors
is very sensitive to long exposure to air and especially to humid
atmosphere. For the Raman measurements a DILOR XY triple spectrometer
combined with a nitrogen cooled CCD detector was used. All Raman data
were obtained at nearly backscattering geometry. The photon excitation was
provided by the 488-nm line of an Ar$^+$ ion laser with laser power equal to
15W/cm$^2$. The estimated additional heating did not exceed 5K.

\section{EXPERIMENTAL RESULTS}

All measurements were performed with the polarization of the incident and
scattered light parallel to the basal plane of the crystal, i.e. the 
CuO$_2$-planes. It was possible to measure the A$_{1g}$, B$_{1g}$, and B$_{2g}$
symmetry components of the Raman scattering cross-section. In addition to the
previously published phonon peaks ($\approx$123cm$^{-1}$,
$\approx$169cm$^{-1}$, $\approx$490cm$^{-1}$, $\approx590$cm$^{-1}$,
$\approx$610cm$^{-1}$) \cite{Gas89,Kall94} we have detected some
additional phonons ($\approx$240cm$^{-1}$, $\approx$300cm$^{-1}$,
$\approx$330cm$^{-1}$, $\approx375$cm$^{-1}$)
which we believe are the defect induced infrared active phonons.  For all
scattering geometries the spectra for temperatures well below T$_c$ were divided
by the spectra just above T$_c$ in order to emphasize the redistribution of
the scattering intensity in the superconducting state compared to the normal
state.
The results of the electronic Raman scattering in the crystals of Tl-2201
(T$_c$=80K,90K) are shown on Figs. \ref{f1}-\ref{f5}.
In the crystal with
T$_c$=80K the B$_{1g}$ scattering component measured in X'Y' configuration
shows a well-defined peak at 430$\pm$15 cm$^{-1}$ (Fig. 1a). The X' and Y'
axes are rotated by 45$^\circ$ with respect to the crystal X and Y axes,
respectively, which are parallel to the crystallographic axes. The B$_{2g}$
scattering component in Fig. \ref{f1}b is less intense, but shows also a broad
maximum with an average frequency of 380$\pm$35cm$^{-1}$. Raman spectra in the
X'X' and XX geometries are presented in Fig. \ref{f2}a and b, showing spectra
of A$_{1g}$ + B$_{2g}$ and A$_{1g}$ + B$_{1g}$ scattering components,
respectively. In order to evaluate the A$_{1g}$ scattering component we
subtracted the B$_{1g}$ and B$_{2g}$ components (see Fig. \ref{f1}a,b) from
the XX and X'X' spectra, respectively. As one can see from Fig. \ref{f3}a,b
the A$_{1g}$ scattering component peaks in both cases, at 345$\pm$20cm$^{-1}$. For
the crystal with T$_c$=90K we found peaks of the B$_{1g}$, A$_{1g}$ and
B$_{2g}$ scattering components at 460$\pm$15cm$^{-1}$,
350$\pm$20cm$^{-1}$, and 400$\pm$35cm$^{-1}$, respectively
(Fig. \ref{f4}, lower panel).

Another very important observation is that the low frequency behavior of
the electronic Raman scattering exhibits strong anisotropy with respect
to the symmetry components. One can see in Fig.4a (upper and lower panel)
that the intensity decrease of the B$_{1g}$ scattering component toward lower
frequencies fits the $\omega^3$ law predicted by Devereaux et al. \cite{Dev95}.
For the A$_{1g}$ and B$_{2g}$ scattering components in Fig. \ref{f4}b and c,
respectively, there is a linear decrease, which also agrees with the predictions
by Deveraux et al. \cite{Dev95}.
A summary of our results on Tl-2201 is presented in Table
\ref{t1}. 

In order to follow the temperature behavior of the superconductor gap,
we have measured the temperature dependence of the electronic Raman
scattering. Following Devereaux et. al.\cite{Dev94,Dev95}, we assume that 
the peak in the B$_{1g}$ component of the electronic scattering corresponds
to the value of 2$\Delta_{max}$. In Fig. \ref{f5}a and b, respectively, we
show the B$_{1g}$ and A$_{1g}$ scattering component of Tl-2201 (T$_c$=80K) at
different temperatures between 10K and T$_c$ divided by the spectrum at 100K.
The experiments for the 90-K crystal yielded a similar behavior.

With increasing temperature the intensity of the peak in Fig. \ref{f5}a
associated with the pair breaking process decreases and the maximum shifts
slightly to lower frequencies. Obviously, the temperature dependence of the
superconductor gap does not follow the BCS behavior.
In other words, upon cooling below T$_c$ the gap opens more abruptly than
predicted by BCS theory. These results are similar to results reported
for underdoped Bi-2212 \cite{Hof95}. Because the peak position of the A$_{1g}$
scattering component in Fig. \ref{f5}b has larger error bars compared to the
B$_{1g}$ component, one cannot definitely say whether the data fit the BCS
behavior or not.

We also searched for superconductivity-induced changes in frequency and
linewidth of the optical phonons. With the resolution of 1cm$^{-1}$ we have
not observed such changes. Upon heating from 10K up to 200K the frequencies
of all phonons decreased and the linewidths increased monotonically. 

\section {DISCUSSION}
The Raman scattering intensity can be written in terms of the
differential scattering cross section\cite{Dev95}:
\begin{equation}
\frac{\partial^2\sigma}{\partial\omega\partial\Omega}=
\frac{\omega_s}{\omega_i}r_0^2S_{\gamma\gamma}(\vec{q},\omega)
\end{equation}
with
\begin{equation}
S_{\gamma\gamma}(\vec{q},\omega)=-\frac{1}{\pi}\left[1+n(\omega)\right]\Im
m\chi_{\gamma\gamma}(\vec{q},\omega)
\end{equation}
Here r$_0=e^2/mc^2$ is the Thomson radius, $\omega_i (\omega_s)$ is the
frequency of incident (scattered) photon, $\hbar$ and k$_B$ were set to 1.
S$_{\gamma\gamma}$ is the generalized structure function, which is connected
to the imaginary part of the Raman response function $\chi_{\gamma\gamma}$
through the fluctuation-dissipation theorem; 
$n(\omega)=1/[\exp (\omega/T)-1]$ is the Bose-Einstein distribution function.
The Raman response function can be written as \cite{Kle84}:
\begin{equation}\label{rares}
\chi_{\gamma\gamma}(\vec{q},\omega)=\langle\gamma^2_{\vec{k}}\lambda_{\vec{k}}
\rangle -\frac{\langle\gamma_{\vec{k}}\lambda_{\vec{k}}\rangle^2}
{\langle\lambda_{\vec{k}}\rangle}
\end{equation}
with the Raman vertex $\gamma_{\vec{k}}$ written as 
\begin{equation}
\gamma_{\vec{k}}(\omega_i,\omega_s)=\sum_L\gamma_L(\omega_i,\omega_s)
\Phi_L(\vec{k}),
\end{equation}
where $\Phi_L(\vec{k})$ are either Brillouin zone or Fermi surface harmonics
\cite{Dev95} which transform according to point group transformations of the
crystal and $\lambda_{\vec{k}}$ is the Tsuneto function:
\begin{equation}
\lambda_{\vec{k}}\propto\frac{\left|\Delta_{\vec{k}}\right|^2}
{\omega\sqrt{\omega^2-4\left|\Delta_{\vec{k}}\right|^2}}.
\end{equation}
The brackets $\langle\cdots\rangle$ in Eq. \ref{rares} denote an average
of the momentum $\vec{k}$ over the Fermi surface.

As is obvious, Raman scattering probes only $\left|\Delta\right|^2$. 
Therfore it is not possible to determine whether the gap function changes sign
for different directions of $\vec{k}=(k_x,k_y)$ or not. But nevertheless
the symmetry of the gap function can be inferred from the specific spectral
features of each symmetry component of the electronic Raman scattering.

For the gap of d-wave symmetry ($\Delta_{\vec{k}}=\Delta_{max}\cdot\cos
2\phi$, where $\phi$ is an angle between $\vec{k}$ and the a-axis),
calculations \cite{Dev94,Dev95} predict different low frequency
behavior for the different symmetry components.
For B$_{2g}$ and A$_{1g}$ scattering components it should show a linear
dependence in $\omega$, but for B$_{1g}$ it should be $\sim\omega^3$. The
appearance of a power law in the low frequency scattering characterizes an
energy gap which vanishes on lines on the Fermi surface. An appearance of
$\omega^3$ law in B$_{1g}$ scattering component is specific for
d$_{x^2-y^2}$-wave pairing \cite{Dev95}. The predicted values of the 
peak maxima for the B$_{1g}$, B$_{2g}$ and A$_{1g}$ scattering components are
$\sim 2\Delta_{max}$,  $\sim 1.6\Delta_{max}$ and $\sim 1.2\Delta_{max}$, 
respectively. These above mentioned peculiarities appear in our data. Indeed,
the low frequency behavior of the B$_{1g}$ scattering component
definitely differs from a linear behavior as seen in Fig. \ref{f4}a, whereas
for the A$_{1g}$ and B$_{2g}$ scattering components it is linear in $\omega$
(see Fig. \ref{f4}b,c). For both crystals, the B$_{1g}$ scattering component
peaks at a higher frequency than the  B$_{2g}$, which in turn peaks at a higher
frequency than the A$_{1g}$ component. 

Since Raman scattering does not probe the phase of the order parameter it is
important to take into consideration other types of the pairing which can
also have nodes on the Fermi surface,  but do not change the sign, i.e.
$s+id$-pairing, or strongly anisotropic s-pairing.
For the $s+id$-pairing \cite{Dev95}
$(\Delta (k)=\Delta_s + i\Delta_d\cos 2\phi)$ one gets the threshold at
$\omega = 2\Delta_s$ (minimum pair breaking energy). While A$_{1g}$ and
B$_{2g}$ scattering components exhibit a jump at this frequency, the
B$_{1g}$ scattering component shows a continuous rise from zero and up to the
peak at $\omega = 2\Delta_{max} =2\sqrt{(\Delta_s^2+\Delta_d^2)}$. The A$_{1g}$
and B$_{2g}$ scattering components also show broad maxima as in the case of
pure d$_{x^2-y^2}$-wave pairing, but these maxima will be cut-off toward lower
frequencies due to the strong jump at 2$\Delta_s$. Thus one should observe a
low-frequency cutt-off in both A$_{1g}$ and B$_{2g}$ scattering components,
which, however, is not observed in our data.

For anisotropic s-pairing, showing the minimum of the gap on the diagonals
of the two-dimensional Brillouin zone, ($\Delta (k) = \Delta_0 +
\Delta_1\cos^42\phi$) one gets a single threshold on 2$\Delta_0$ for all
scattering components as well as a peak at $\omega = 2\Delta_{max} =
2(\Delta_0 + \Delta_1)$ for the B$_{1g}$ scattering component. Therefore we
will expect a picture which is very similar to the case of $s+id$-pairing,
with one exception. The B$_{1g}$ scattering component should show an
additional shoulder at the same position where the A$_{1g}$ and B$_{2g}$
scattering components show peaks \cite{Dev95}. This is also not the case for
our data. In principle one can assume $\Delta_0$ to be very small or even
zero. In this case one gets peaks at 2$\Delta_{max}$, 0.6$\Delta_{max}$ and
0.2$\Delta_{max}$ for the B$_{1g}$, B$_{2g}$ and A$_{1g}$ scattering
components \cite{Dev95}, respectively. In addition, the low frequency behavior
of the B$_{1g}$ scattering component will be linear. This also contradicts
our results.

Recently the model calculations of Devereaux et al. were criticized by Krantz
and Cardona \cite{Kra94,Kra95}. The main argument against this theoretical
model is that the realistic electronic band structure of the crystal is
important, but that the one-sheeted Fermi surface approximation used by
Devereaux et al.\cite{Dev95} is inappropriate. The authors of
Ref.\onlinecite{Kra95} used a numerical model based on the LDA band
structure calculations for YBaCuO in order to take into account the
multisheeted Fermi surface of the superconductors with several CuO$_2$ -planes.
It was pointed out that for the $\Delta = \Delta_0 cos 2\phi$ (d-wave pairing)
and a multisheeted Fermi surface the calculations lead to a contradiction
with the experiment, i.e. the A$_{1g}$ and B$_{1g}$ scattering components
peak at the same position 2$\Delta_0$. In order to get consistency with the
experiment, different types of the order parameter on different sheets of the
Fermi surface were proposed. Only in this case the calculations in
Ref.\onlinecite{Kra95} were able to get different positions of the maxima of
the B$_{1g}$, A$_{1g}$, and B$_{2g}$ scattering components. For a one-sheeted
Fermi surface the authors of Ref.\onlinecite{Kra95} found identical positions
of the maxima for the A$_{1g}$ and B$_{1g}$ components, but a different
position for the B$_{2g}$ component. Hence it was concluded that any
difference in peak position of the A$_{1g}$ and B$_{1g}$ component is only
consistent with multiband scattering of a  multisheeted Fermi surface and
different gap symmetries for each of the sheets. For superconductors with
one CuO$_2$-plane, a multisheeted Fermi
surface is invoked originating from Tl-like s-states \cite{Kra95}
(Tl-2201) or from Sr-doping\cite{Kra} (La$_{2-x}$Sr$_x$CuO$_4$) in
order to yield a difference in peak position for the A$_{1g}$ and B$_{1g}$
components. However, no experimental proof for such a Fermi surface
contribution exists so far. Moreover, the calculations in
Ref.\onlinecite{Kra95}
failed in explaining of the symmetry-dependent low-frequency dependence of the
Raman scattering intensity, whereas this important experimental fact was
observed not only in our experiments, but also in Bi-Sr-Ca-Cu-O
\cite{Sta92,Hof95,Dev94}, Y-Ba-Cu-O\cite{Hac88,Coo88} and
La-Sr-Cu-O \cite{Che94} systems. In addition, it is obvious that all
superconductors with different crystal structures have a different
electronic structure. Hence, if the multiband scattering model would be
crucial we would expect absolutely different behavior for the different
superconductors which is actually in contradiction with existing experimental
results. Even if one compares the superconductors with the same number of
CuO$_2$ planes,
one finds that, while the interplanar distance (distance between Cu atoms in
different CuO$_2$-planes) is quite similar, the dimpling (in-plane Cu-O-Cu
angle) differs very much from compound to compound (see Table \ref{t2}).
YBa$_2$Cu$_3$O$_7$ exhibits the largest dimpling compared to other compounds.
Strong dimpling should lift the degeneracy of otherwise identical
CuO$_2$-planes or Fermi surface sheets. This dimpling can strongly affect the
LDA calculations because the interplanar interaction should depend on this
parameter.

And finally on top of that, use of the effective mass approach is very much
questionable in case of high temperature superconductors, because
this approach can be used only for nonresonant Raman scattering \cite{Abr}.
In high temperature superconductors we, however, are always in the regime of 
the resonant scattering. Moreover, the electron correlation effects in HTSC are
not treated sufficiently by LDA.

In contrast to the conclusion of Ref. \onlinecite{Kra95} our experiments show that the
one-CuO$_2$-plane compound Tl-2201 shows very similar behavior compared to
compounds with several CuO$_2$-planes, such as Tl-2223, Bi-2212 and YBaCuO
\cite{Hac88,Coo88,Hof94,Sta92,Hof95,Nem93,Che92,Che94,Dev94,Dev95,Kra94,Kra95}.
We also found that the frequency of the B$_{1g}$ maximum scales
with T$_c$, and it corresponds to the value 2$\Delta_{max}/k_BT_c=7.6\pm 0.4$.
This value is  very close to the values (with the exception of
Nd$_{2-x}$Ce$_x$CuO$_4$ \cite{Hac}) found for other high temperature
superconductors as shown in Table \ref{t3}.

The temperature dependence of the gap (B$_{1g}$ component in Fig. \ref{f5}a)
in our experiment differs from the BCS behavior, i.e. upon cooling the gap
opens more abruptly than predicted by BCS theory.
This is consistent with the spin fluctuation theory of high temperature
superconductivity \cite{Mon92}, favoring d$_{x^2-y^2}$-wave pairing. The
model considers pair binding as well as pair breaking effects due to the
spin fluctuations. Gap opening leads to a suppression of low-frequency spin
fluctuations and therefore to reduced pair-breaking. Therefore in underdoped
crystals (we consider our Tl-2201 crystals as underdoped) this effect
will lead to a more abrupt opening of the gap upon cooling below T$_c$ 
compared to BCS behavior.

In conclusion, we presented measurements of the electronic Raman scattering on
high-T$_c$ Tl-2201 single crystals with one CuO$_2$-plane per unit cell.
The peculiarities of the electronic Raman scattering, i.e. the power law
frequency dependence of the diferent scattering components at low frequencies,
their different peak positions as well as the values of
2$\Delta_{max}/k_BT_c=7.6\pm 0.4$ are found to be very similar in compounds
with one and several CuO$_2$ planes.
All nearly optimally doped high-T$_c$ superconductors (with the exception of
(Nd,Ce)-214 \cite{Hac}) show a very similar behavior of the electronic Raman
scattering consistent with the d$_{x^2-y^2}$- wave symmetry of the underlying 
order parameter.

\section {ACKNOWLEDGMENTS}
This work was supported by DFG through SFB 341 and by BMBF FKZ 13 N 6586.
One of the authors L.V.G acknowledges support from the Alexander von Humboldt
Foundation and expresses his gratitude for the hospitality at the
2.Physikalisches Institut RWTH-Aachen.

\newpage
\begin{figure}
\caption{Electronic Raman scattering of Tl$_2$Ba$_2$CuO$_{6+\delta}$
(T$_c$=80K).
Shown are spectra at T=10K and 100K, and the divided spectra
I(T=10K)/I(T=100K) for a) B$_{1g}$ and b) B$_{2g}$ scattering components.
The phonon at$\sim$490cm$^{-1}$ is due to the leakage of the A$_{1g}$
scattering component while the other phonons are deffect induced infrared
activ phonons. Note that the peak in the divided spektrum does not coincide
with the phonon at $\sim$490cm$^{-1}$. The light polarization is shown in
relation to the crystal axes.}
\label{f1}
\end{figure}

\begin{figure}
\caption{Electronic Raman scattering of Tl$_2$Ba$_2$CuO$_{6+\delta}$
(T$_c$=80K) in a) A$_{1g}$+B$_{2g}$ (XX) and b) A$_{1g}$+B$_{1g}$ (X'X') 
scattering geometries. Shown are spectra at T=10 K and 100 K, and divided
spectra I(T=10K)/I(T=100K). The phonon at $\sim$490cm$^{-1}$  was cut off in
order to show the variations of electronic Raman scattering.
The polarization is shown in
relation to the crystal axes.}
\label{f2}
\end{figure}

\begin{figure}
\caption{Electronic Raman scattering of Tl$_2$Ba$_2$CuO$_{6+\delta}$ in
A$_{1g}$ scattering geometry evaluated from a) XX and b) X'X' spectra. Shown
are spectra at T=10 K and 100 K, and divided spectra I(T=10K)/I(T=100K).
The phonon at $\sim$490cm$^{-1}$ was cut off in order to show the changes of 
electronic Raman scattering.}
\label{f3}
\end{figure}

\begin{figure}
\caption{Electronic Raman scattering in Tl-2201 with T$_c$=80 (upper panel)
and 90K (lower panel). Shown
are divided spectra I(T=10K)/I(T=100K) for a) B$_{1g}$, b) A$_{1g}$ and c)
B$_{2g}$ scattering components.}
\label{f4}
\end{figure}

\begin{figure}
\caption{Temperature dependence of the electronic Raman scattering. Shown are
spectra for 10, 35 and 50K (upper panel), and values of $\Delta(T)/\Delta_0$
(lower panel) evaluated for a) B$_{1g}$ and b) A$_{1g}$ scattering components.
In the lower panel the temperature dependence of the peak positions is
compared to that of the BCS theory}
\label{f5}
\end{figure}
\newpage
\begin{table}
\caption{Peak positions of the different scattering components of electronic
Raman scattering. The reduced gap values 2$\Delta/k_BT_c$ are referred to the
B$_{1g}$ peak position, n is the number of CuO$_2$-planes per unit cell.}
\begin{tabular}{ccccccc}
Compound & n & T$_c$ [K] & B$_{1g}$ [cm$^{-1}$] & A$_{1g}$ [cm$^{-1}$] &
B$_{2g}$[cm$^{-1}$] & 2$\Delta/k_BT_c$\\
\tableline
Tl$_2$Ba$_2$CuO$_6$ & 1 & 90 & 460$\pm$15 & 350$\pm$20 & 400$\pm$35
& 7.4$\pm$0.4\\
Tl$_2$Ba$_2$CuO$_6$ & 1 & 80 & 430$\pm$15 & 345$\pm$35 & 380$\pm$35 &
7.8$\pm$0.4\\
\end{tabular}
\label{t1}
\end{table}

\begin{table}
\caption{Interplanar Cu-Cu distance and dimpling (Cu-O-Cu angle in
CuO$_2$-plane)
for different high temperature superconductors with two (n=2) CuO$_2$ planes
per unit cell.}
\begin{tabular}{ccccc}
Compound & n & T$_c$ [K] & interplane distance [{\AA}] & dimpling [deg]\\
\tableline
YBa$_2$Cu$_3$O$_{7-\delta}$ \cite{Jor90} & 2 & 92 & 3.37 & 164\\
TlBa$_2$CaCu$_2$O \cite{Mor88} & 2 & 103 & 3.20 & 177\\
Tl$_2$Ba$_2$CaCu$_2$O$_8$ \cite{Sub88} & 2 & 110 & 3.17 & 178\\
Bi$_2$Sr$_2$CaCuO$_8$ \cite{Bor88} & 2 & 84 & 3.44 & 179\\
La$_{1.6}$Sr$_{0.4}$CaCu$_2$O$_{5.94}$ \cite{Cav90} & 2 & 55 & 3.40 & 175\\
\end{tabular}
\label{t2}
\end{table}

\begin{table}
\caption{Peak positions of the different symmetry components and reduced gap
 values 2$\Delta/k_BT_c$ for different investigated high temperature 
superconductors with different number n of CuO$_2$ planes per unit cell.}
\begin{tabular}{cccccc}
Compound & n & T$_c$ [K] & B$_{1g}$ [cm$^{-1}$] & A$_{1g}$ [cm$^{-1}$]
& 2$\Delta/k_BT_c$\\ \tableline
YBa$_2$Cu$_3$O$_{7-\delta}$ \cite{Che92} & 2 & 89.7 & 420 & 310 & 7.6\\
 & & 93.7 & 550 & 310 & 8.4\\
Bi$_2$Sr$_2$CaCuO$_8$ \cite{Hof95} & 2 & 81 & 460 & 280 & 8.2\\
 & & 86 & 520 & 330 & 8.7\\
La$_{2-x}$Sr$_x$CuO$_4$ \cite{Che94} & 1 & 37 & 200 & 125 & 7.8\\
Tl$_2$Ba$_2$Ca$_2$Cu$_3$O$_{10}$ \cite{Hof94} & 3 & 118 & 610 & 430 & 7.5\\
Nd$_{2-x}$Ce$_x$CuO$_4$ \cite{Hac} & 1 & 19.3 & 70 & 70 & 5.2\\
\end{tabular}
\label{t3}
\end{table}

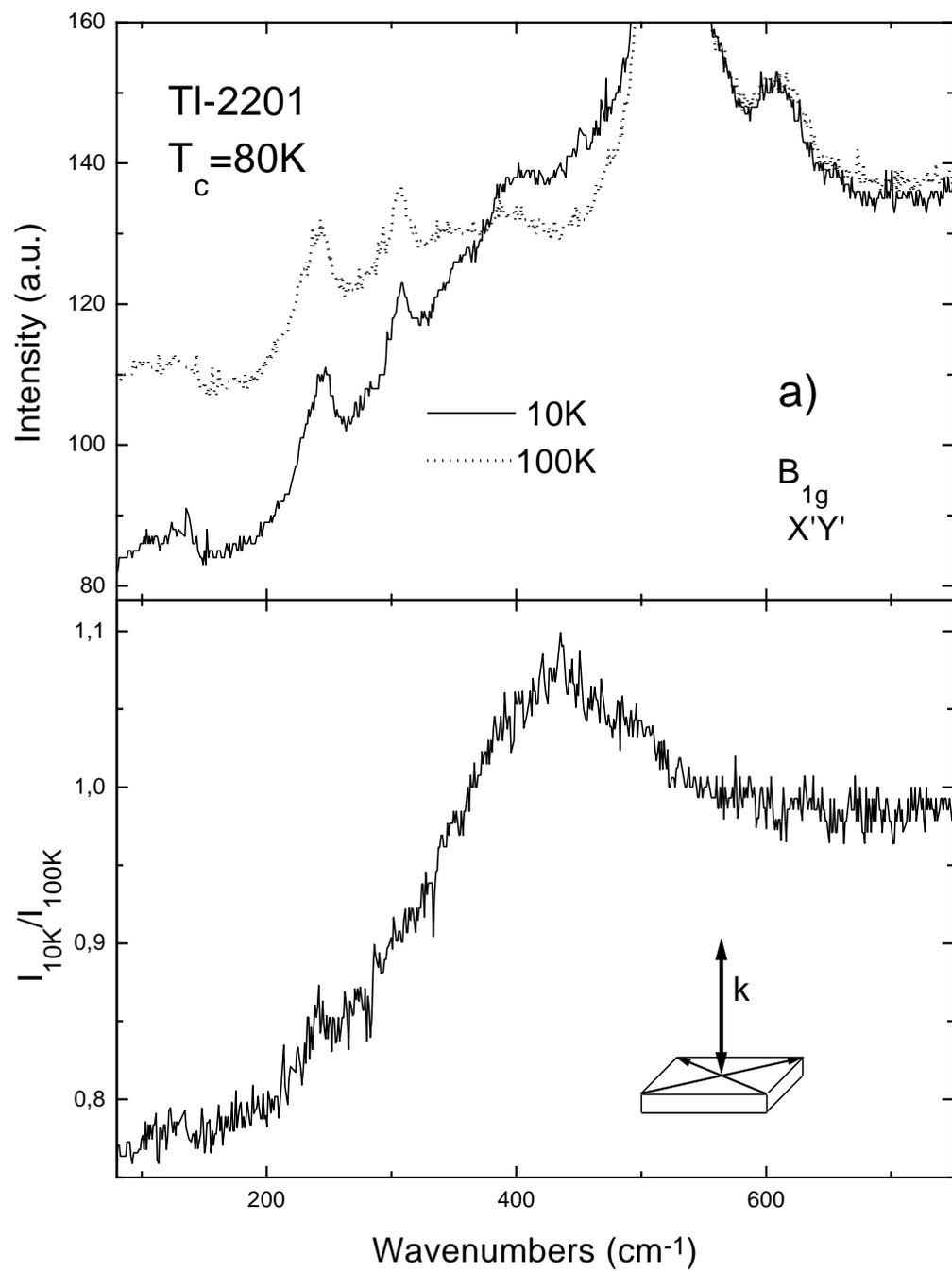
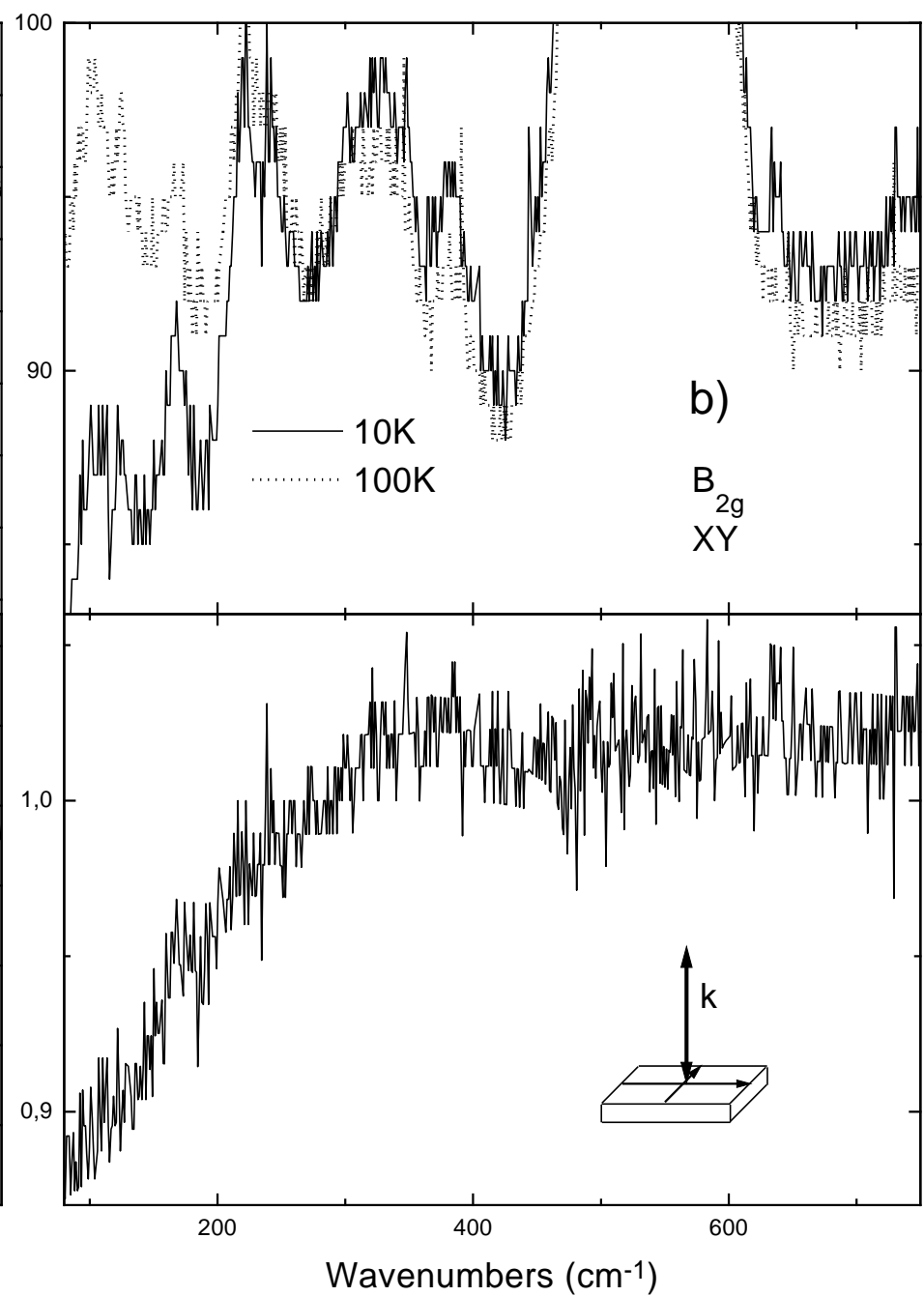
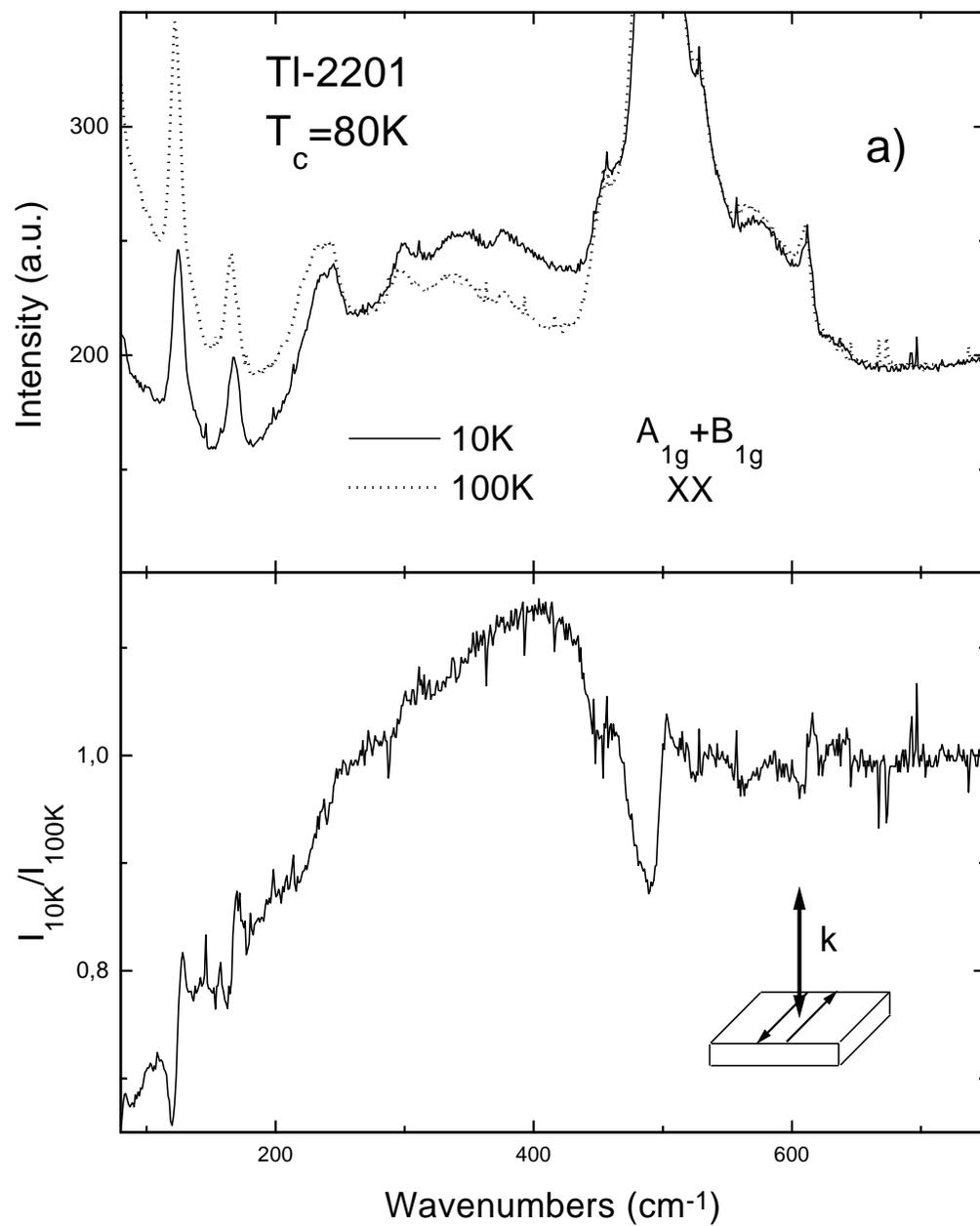
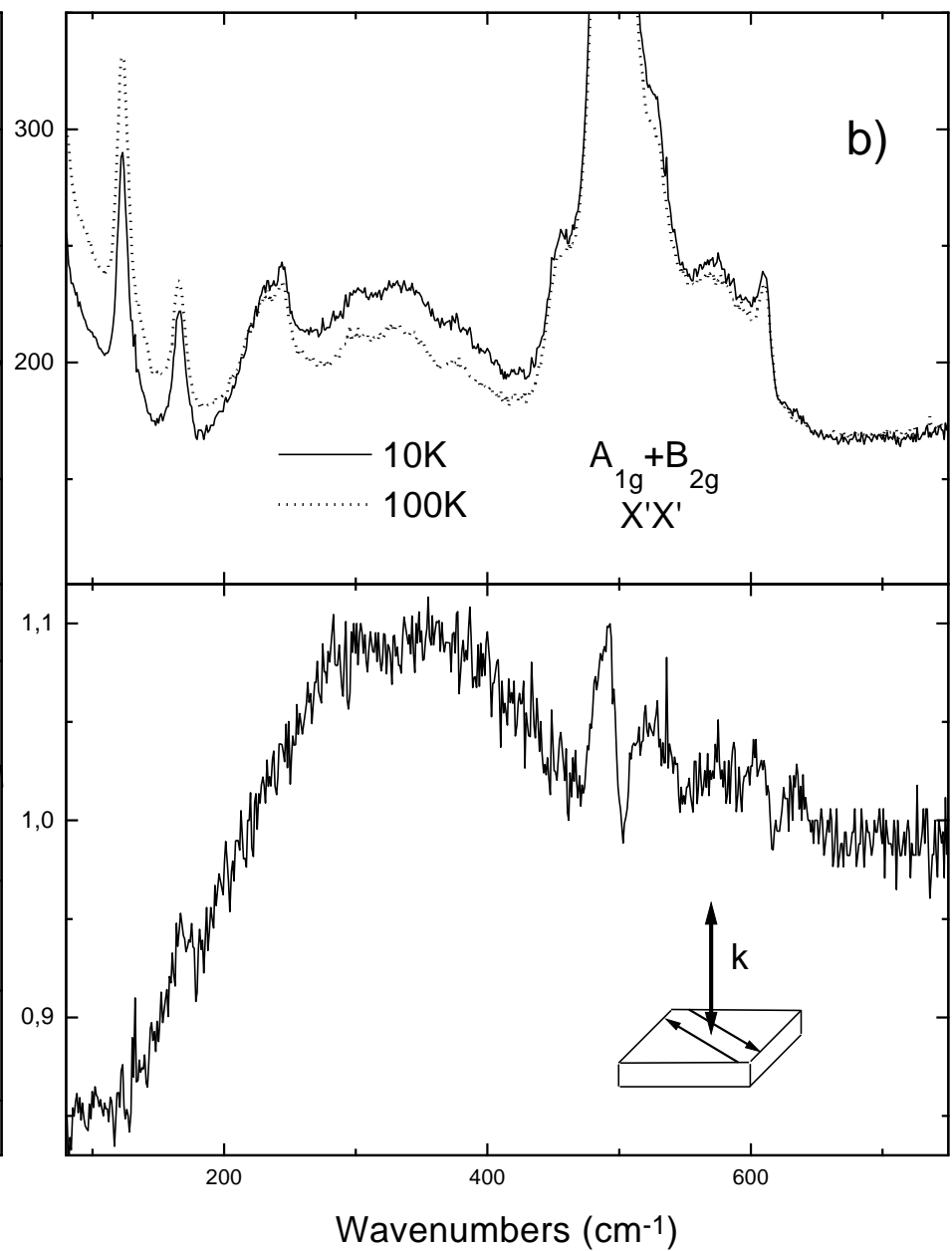
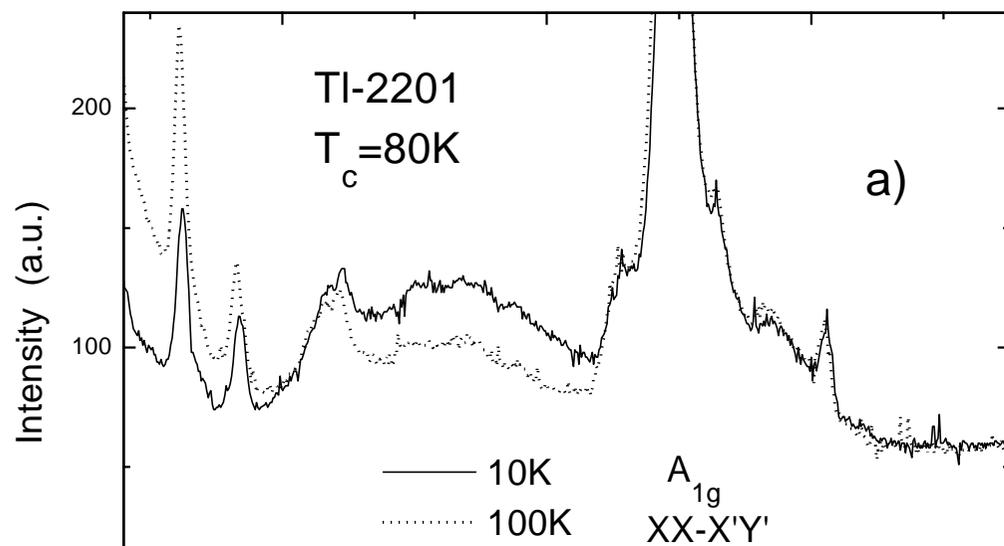
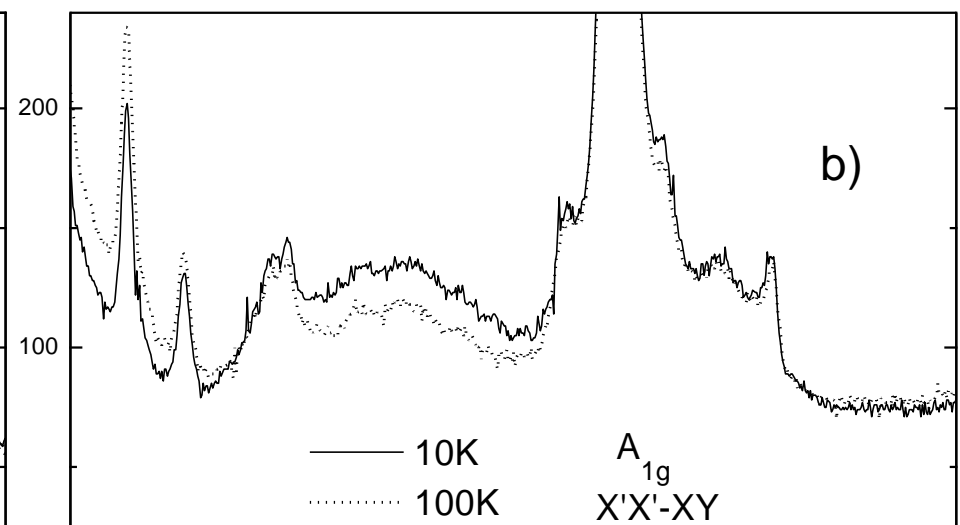
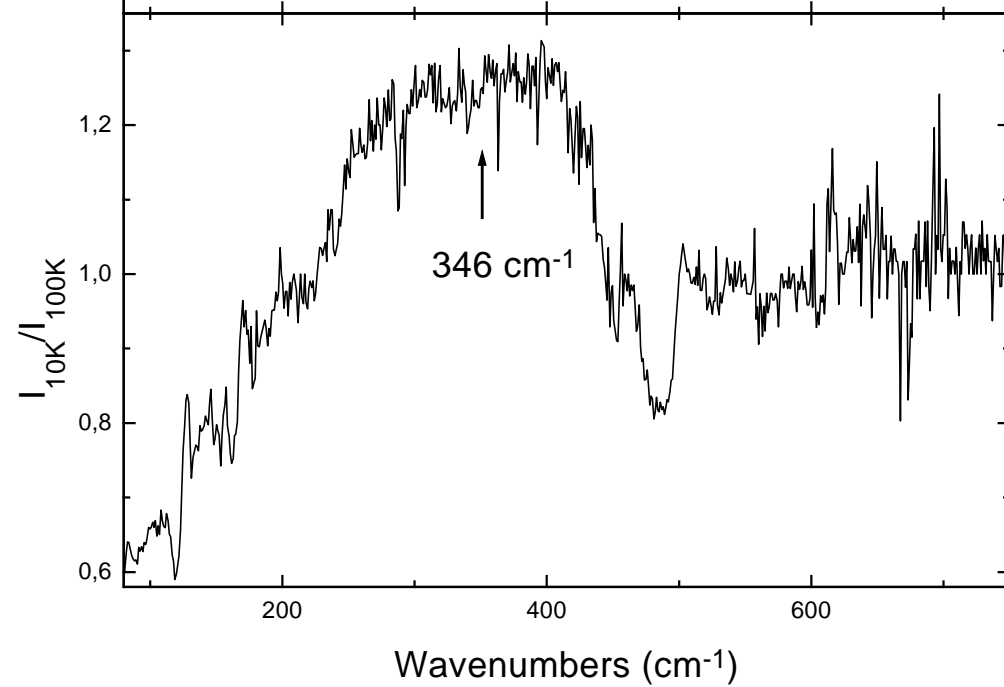
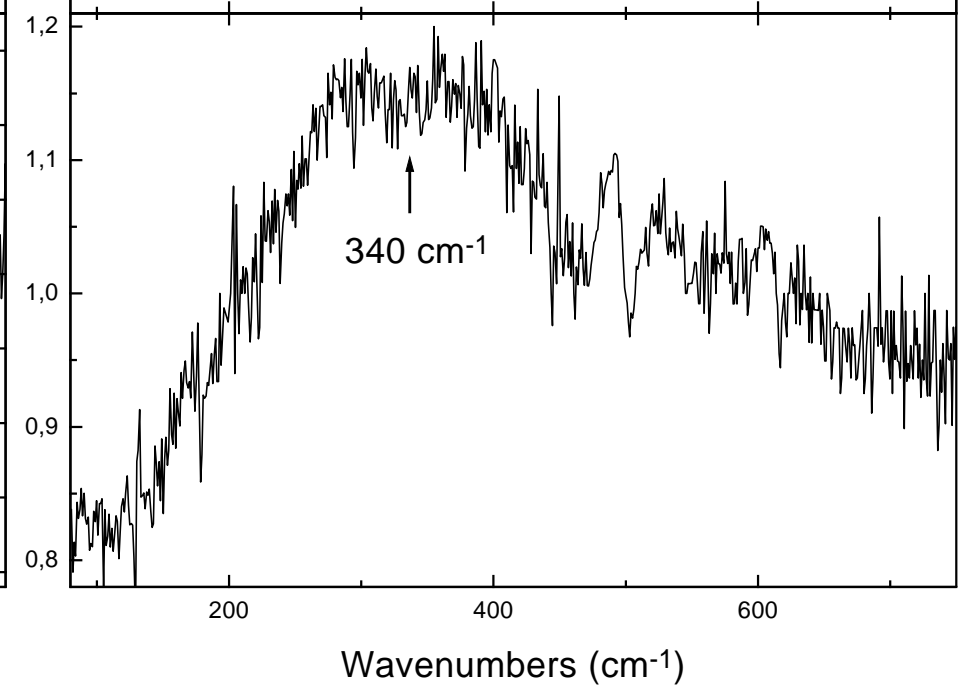
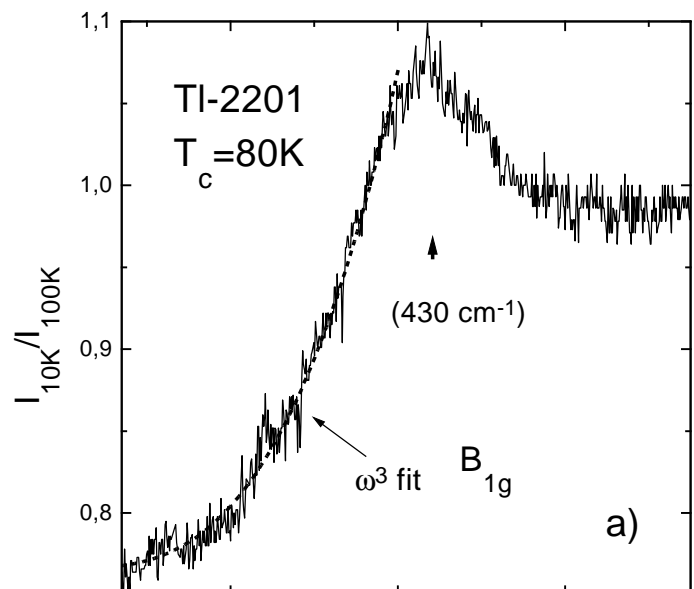
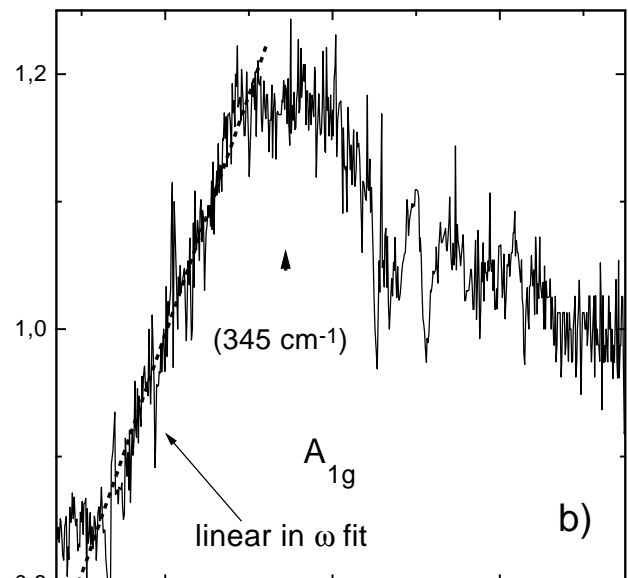
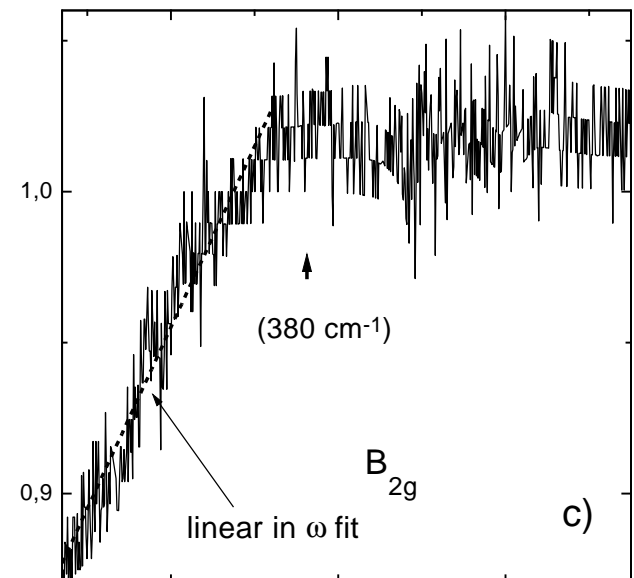
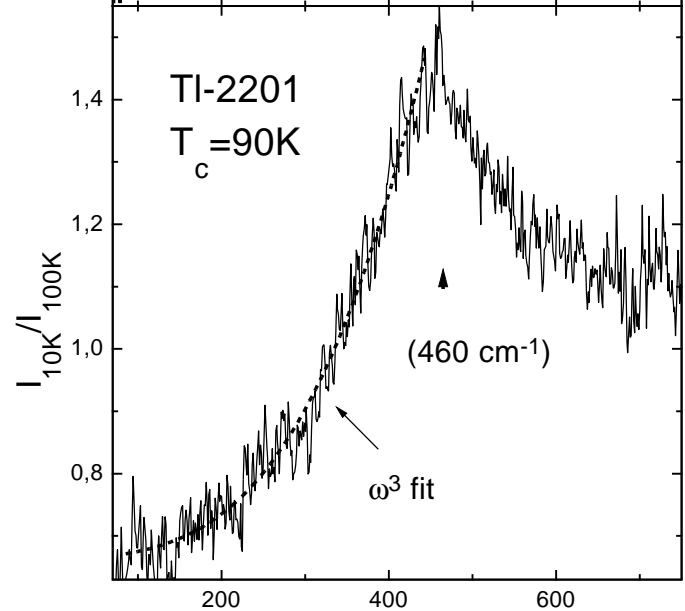
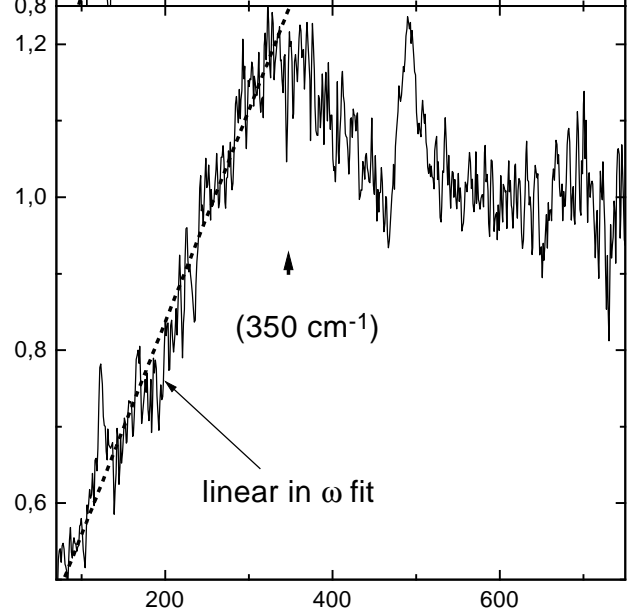
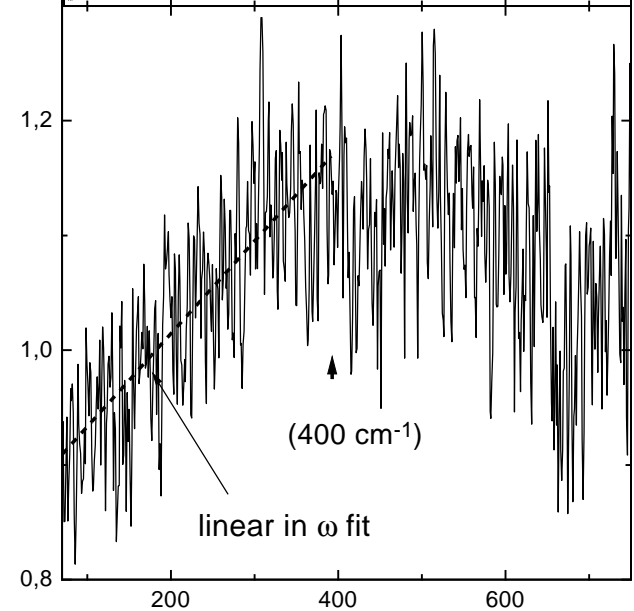
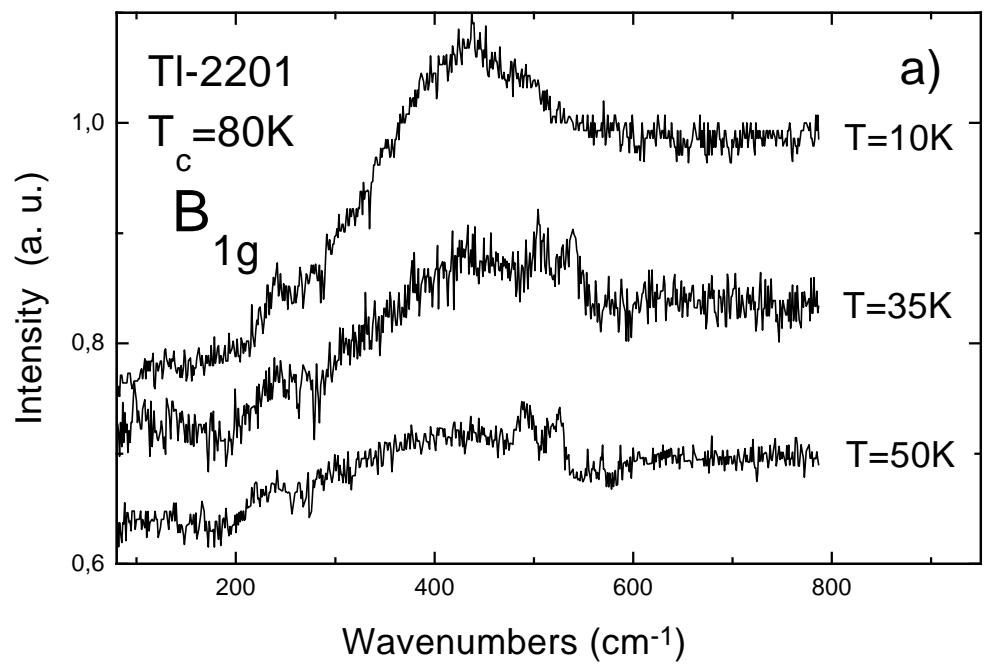
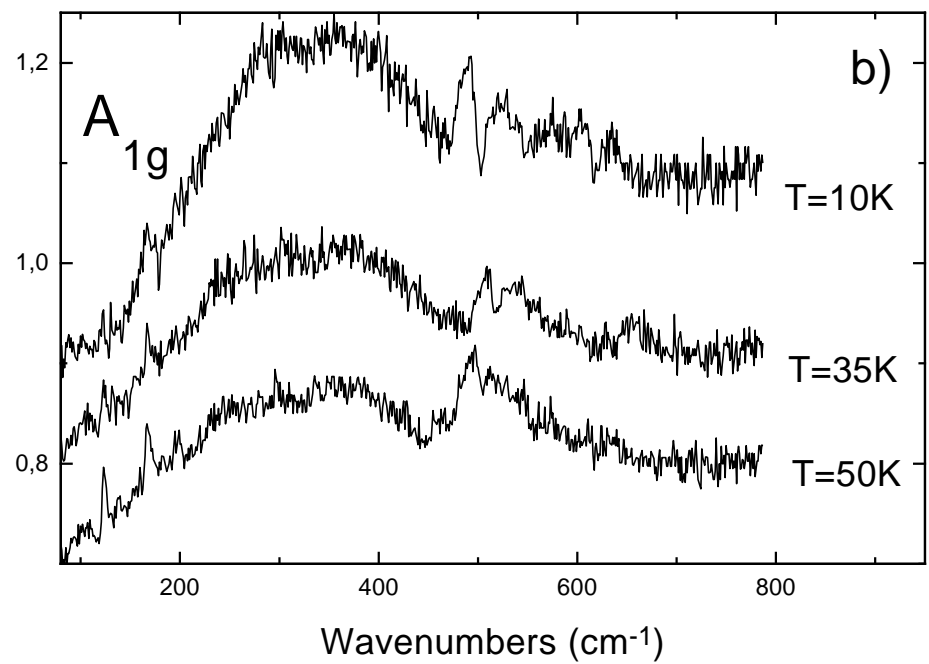
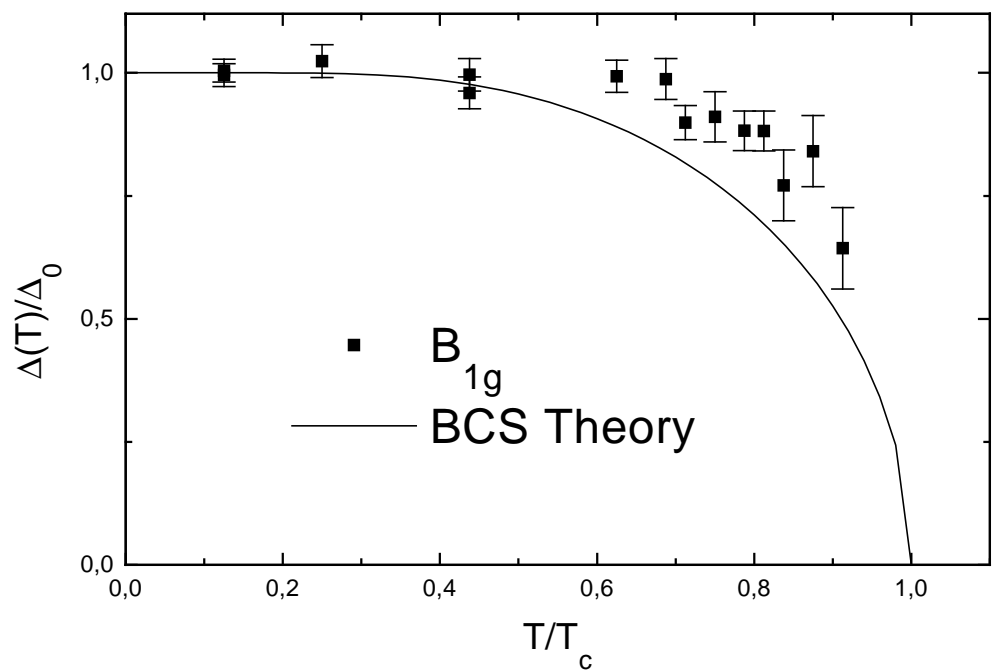
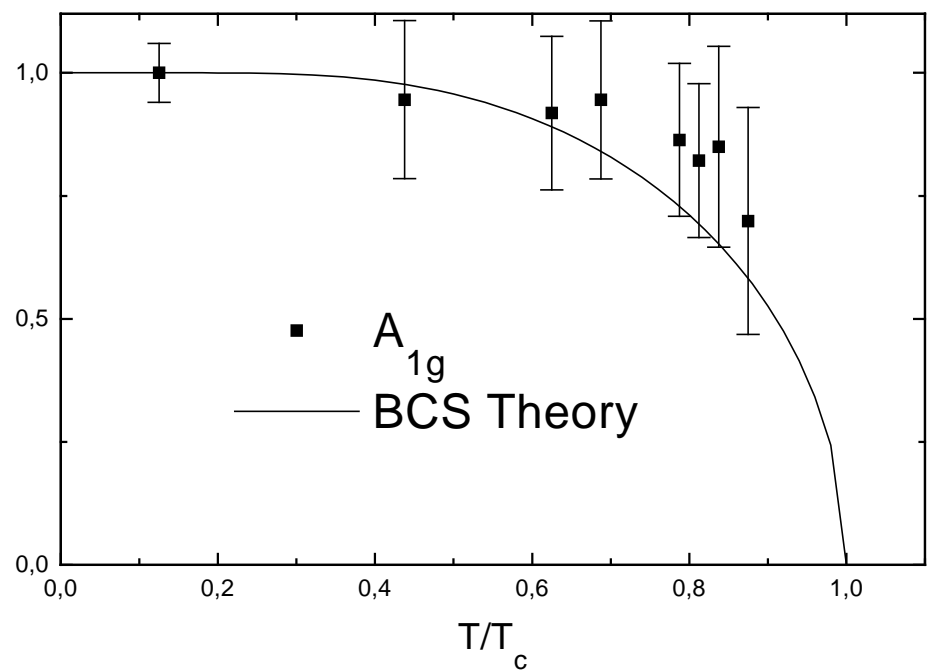
\end{document}